\documentclass[aps,showpacs,preprintnumbers,amsmath, amssymb]{revtex4}

\usepackage{float}
\usepackage{graphics,epsfig}
\usepackage{graphicx}
\usepackage{dcolumn}
\usepackage{bm}

\begin{document}
\baselineskip=0.8 cm
\title{{\bf Phase transition in the holographic model of superfluidity with backreactions}}

\author{Yan Peng$^{1}$, Xiao-Mei Kuang$^{2}$, Yunqi Liu$^{1}$, Bin Wang$^{2}$}
\affiliation{$^{1}$ Department of Physics, Fudan
University, Shanghai 200433, China\\
$^{2}$ INPAC, Department of Physics and Shanghai
Key Lab for Particle Physics and Cosmology,
Shanghai Jiao Tong University, Shanghai 200240,
China}

\vspace*{0.2cm}
\begin{abstract}
\baselineskip=0.6 cm
\begin{center}
{\bf Abstract}
\end{center}

In the fully back reacted geometry we develop a
supercurrent solution which corresponds to a
deformation of superconducting black holes by the
spatial component of gauge fields with a
non-trivial radial dependence. We investigate the
influence of the backreaction on the condensation
and the phase structure in the AdS black hole
spacetime. Different from that observed in the
probe limit, we find that two operators cannot
give consistent information on the phase diagram.
We argue that only one of these two operators can
reflect the real property of the condensation and
phase structure in the holographic
superconductor. We also generalize our study to
the associated phase diagram in the AdS soliton
background.

\end{abstract}

\maketitle
\newpage
\vspace*{0.2cm}

\section{Introduction}

The AdS/CFT correspondence \cite{Maldacena,S.S.Gubser-1,E.Witten}
has provided us a novel way to describe the strongly coupled field
theories in a weakly coupled gravitational system. In recent years,
it has been found that the gauge gravity duality can be used to
provide some insights into superconductivity \cite{S.A.
Hartnoll,C.P. Herzog,G.T. Horowitz-1}. There exists a gravitational
system which closely mimics the behavior of a superconductor. When
the temperature of a black hole drops below the critical value, the
bulk AdS black hole becomes unstable and scalar hair condenses on
the black hole background. The instability of the bulk black hole
corresponds to a second order phase transition from normal state to
superconducting state and the emergence of the scalar hair in the
bulk AdS black hole corresponds to the formation of a charged
condensation in the boundary dual CFTs. The gravity models with the
properties of holographic superconductors have attracted
considerable interest for their potential applications to the
condensed matter physics, see for examples \cite{G.T.
Horowitz-2}-\cite{Gary T}.

Recently, the investigation has been further generalized by
presenting a DC supercurrent type solution \cite{C. Herzog-2,P.
Basu}. In terms of AdS/CFT correspondence, the supercurrent states
correspond to a deformation of superconducting black holes by the
spatial component of the gauge fields with a non-trivial radial
dependence. It has been shown that with the introduction of a
chemical potential for supercurrent, the critical temperature of the
superconducting transition decreases and furthermore at some point
the order of phase transition changes from second order to first
order. The novel phase diagram brought by the supercurrent is
interesting. It further enriches the phase structure observed in the
St$\ddot{u}$ckelberg mechanism \cite{S. Franco,Q.Y. Pan}.

In \cite{C. Herzog-2,P. Basu} the supercurrent solution was obtained
within the approximation scheme of neglecting gravity backreaction.
It would be of great interest to examine how the solution changes as
we incorporate the gravity backreaction, especially to investigate
whether the phase diagram can be modified in the fully back reacted
geometry. In the p-wave superfluids system, it was argued that the
order of the phase transition depends on the backreaction \cite{M.
Ammon}. The effects of the backreaction on the order of
superconducting transition have also been observed in \cite{Gary T,
Yunqi Liu, Peng Yan}.  Here we will further examine how the phase
diagram will be modified due to the backreaction in the supercurrent
solution in the background of the AdS black hole.

In studying the holographic superconductor,  the operator in the
dual theory charged under the $U(1)$ is dual to the scalar field
$\psi$ in the bulk. At the AdS boundary the asymptotic behavior of
the scalar field has the form $\psi \sim
\frac{\psi_-}{r^{\lambda_-}}+\frac{\psi_+}{r^{\lambda_+}}$, where
$\lambda_{\pm}= (d\pm \sqrt{d^2 +4m^2l^2})/2$, $d$ is the dimension
of the bulk space and $l$ the AdS radius. Usually the
normalizability gives the freedom to consider solutions either with
$\psi_- = 0$ or $\psi_+ = 0$.  Depending on the choice of boundary
conditions, we can read off the expectation value of an operator
$<O_-> \sim \psi_-$, or of an operator $<O_+> \sim \psi_+$. In order
to apply the formalism in gravity to study the real condensed matter
physics, one may ask which one of the two possible operators can
really reflect the properties of the real condensation. Some works
have been reported in examining these two operators in describing
the condensations \cite{Yunqi Liu,R. Gregory,Q.Y. Pan,S. A.
Hartnoll-3}. In \cite{Yunqi Liu} it was argued that only one of
these two possible operators can reflect the real property of the
condensation in the holographic superconductor. This argument was
further supported by the investigation in dynamics.  In the probe
approximation in \cite{C. Herzog-2,P. Basu}, it was argued that
similar behaviors in DC superconductivity and phase structure appear
in studying both operators. It is of interest to further examine
these two operators in describing the scalar condensation when the
fully back reacted geometry is taken into account.

In addition to consider the AdS black hole
background, we will also generalize our
discussion to the bulk AdS soliton configuration.
Including the spatial component of the gauge
fields with a non-trivial radial dependence in
AdS soliton background, we will examine the
scalar condensation and the order of the phase
transition between the holographic superconductor
and insulator systems. We will compare the
effects of the spatial component of gauge fields
in the superconducting phase transition in the
AdS soliton and AdS black hole backgrounds.

\section{The condensation in AdS black hole background}

\subsection{Equations of motion and boundary conditions}

The general lagrangian density describing a
$U(1)$ gauge field and a conformally coupled
charged complex  scalar field in the Einstein
gravity background with negative cosmological
constant reads \cite{Y. Brihaye}
\begin{eqnarray}\label{lagrange-1}
L&=&R+\frac{6}{l^{2}}-\gamma(\frac{1}{4}F^{\mu\nu}F_{\mu\nu}+|\nabla_{\mu}\psi-iA_{\mu}\psi|^{2}+m^{2}|\psi|^{2}),
\end{eqnarray}
where $l$ is the AdS radius which will be taken
as unity in the following discussion. $\psi(r)$
is the scalar field and $A_{\mu}$ is the Maxwell
field. In order to consider the possibility of a
DC supercurrent, the vector potential should have
both the time component $A_t$ and a spatial
component $A_x$.  $\gamma$ here is the
backreaction parameter.

We are interested in including the backreaction,
so we use the ansatz of the geometry of the
4-dimensional AdS black hole with the form
\begin{eqnarray}\label{metric}
ds^{2}&=&-r^2B(r)e^{D(r)}dt^{2}+\frac{dr^{2}}{r^2B(r)}+r^{2}(e^{C(r)}dx^{2}+dy^{2})~.
\end{eqnarray}
There exists an event horizon $r_{h}$ when $B(r_{h})=0$ and the
Hawking temperature reads
$T=\frac{r_{h}^2B'(r_{h})e^{D(r_{h})/2}}{4\pi}$. The function $C(r)$
in the metric ansatz is introduced by considering the nonzero $x$
component of the Maxwell field \cite{Amin Akhavana}.

Assuming the matter fields in the forms
\begin{eqnarray}\label{MF}
\psi=\psi(r),~~~~~~~~~A=\phi(r)dt+h(r)dx,
\end{eqnarray}
we can obtain equations of motions
\begin{eqnarray}\label{psi-3}
\psi''+(\frac{4}{r}+\frac{D'}{2}+\frac{C'}{2}+\frac{B'}{B})\psi'+(\frac{\phi^{2}}{r^{4}B^{2}e^{D}}-\frac{h^{2}}{r^{4}Be^{C}}-\frac{m^{2}}{r^{2}B})\psi=0,
\end{eqnarray}
\begin{eqnarray}\label{phi-3}
\phi''+(\frac{2}{r}+\frac{C'}{2}-\frac{D'}{2})\phi'-\frac{2\psi^{2}}{r^{2}B}\phi=0,
\end{eqnarray}
\begin{eqnarray}\label{phi-3}
h''+(\frac{2}{r}+\frac{D'}{2}+\frac{B'}{B}-\frac{C'}{2})h'-\frac{2\psi^{2}}{r^{2}B}h=0,
\end{eqnarray}
\begin{eqnarray}\label{phi-3}
C''+\frac{1}{2}C'^{2}+(\frac{4}{r}+\frac{D'}{2}+\frac{B'}{B})C'+\gamma\frac{h'^{2}}{r^{2}e^{C}}+\gamma\frac{2h^{2}\psi^{2}}{r^{4}Be^{C}}=0,
\end{eqnarray}
\begin{eqnarray}\label{phi-3}
B'(\frac{2}{r}-\frac{C'}{2})-\frac{1}{2}BD'C'+\frac{6}{r^{2}}B-\frac{6}{r^2}+\gamma\frac{\phi'^2}{2r^2e^{D}}+\gamma\frac{m^2\psi^2}{r^2}+\gamma
B\psi'^{2}+
\gamma\frac{\psi^2\phi^2}{r^4Be^{D}}-\gamma\frac{Bh'^2}{2r^{2}e^{C}}-\gamma\frac{h^{2}\psi^2}{r^4e^{C}}=0,
\end{eqnarray}
\begin{eqnarray}\label{phi-3}
D'=\frac{4rC'+r^2C'^2+2r^2C''+\gamma
r(\frac{2h'^{2}}{re^{C}}+4r\psi'^{2}+\frac{4\psi^2\phi^2}{r^3B^2e^{D}})}{r(4+rC')}.
\end{eqnarray}

Considering the symmetry
\begin{eqnarray}\label{symmetry}
r \rightarrow ar,~~~~~~~~(x,y,z,\eta)\rightarrow
~(x,y,z,\eta)/a,~~~~~~~\phi\rightarrow a
\phi,~~~~~~h \rightarrow\ a h,
\end{eqnarray}
we can adjust the solutions to satisfy $r_{h}=1$.
Since the equations are coupled and nonlinear, we
have to numerically integrate these equations
from the horizon out to the infinity.

Examining the behavior of the fields near the
horizon $r_{h}=1$, we find
\begin{eqnarray}\label{boudary}
\psi(r)=a+b(r-r_{h})+c(r-r_{h})^{2}+\cdot\cdot\cdot,\\
\phi(r)=u(r-r_{h})+v(r-r_{h})^{2}+\cdot\cdot\cdot,\\
h(r)=\overline{\alpha}+\overline{\beta}(r-r_{h})+\overline{\gamma}(r-r_{h})^{2}+\cdot\cdot\cdot,\\
B(r)=\overline{u}(r-r_{h})+\cdot\cdot\cdot,\\
D(r)=\overline{\upsilon}+\overline{\omega}(r-r_{h})+\cdot\cdot\cdot,\\
C(r)=\overline{s}+\overline{t}(r-r_{h})+\cdots.
\end{eqnarray}

At the AdS boundary, after choosing $m^{2}=-2$,
the scalar and Maxwell fields behave as
\begin{eqnarray}\label{inf}
\psi=\frac{\psi_{-}}{r}+\frac{\psi_{+}}{r^{2}}+\cdot\cdot\cdot,\
\phi=\mu-\frac{\rho}{r}+\cdot\cdot\cdot, \ \
h=\sigma-\frac{\xi}{r}+\cdot\cdot\cdot.
\end{eqnarray}
The constant coefficients above can be related to
physical quantities in the boundary field theory
using the AdS/CFT dictionary. $\mu$, $\rho$ are
the chemical potential and charge density in the
dual theory respectively. $\xi$ is proportional
to the current density and $\sigma$ is the dual
current source.

For the value of $m^{2}$ we choose, both
$\psi_{-}$ and $\psi_{+}$ are normalized. So
either $\psi_{-}$ or $\psi_{+}$ can be
expectation values of the operators in the field
theory $\psi_i\sim<O_i>$. In this work we will
study both the $\psi_+=0$ case and $\psi_-=0$
case and compare these two operators in the
description of the superfluidity.

\subsection{Effects on the phase transition}

It was shown in \cite{C. Herzog-2,P. Basu} that
there exists a first order transition to the
normal state in the phase structure at low
temperature as the fluid velocity increased. At
temperatures close to the critical value $T_c$,
the transition becomes second order. Their results
were obtained in the probe approximation and they
argued that their results were valid no matter which
operator they choose. Here we will examine
whether the phase diagram will be modified in the
fully back reacted geometry. We will examine the
effects of different operators on the
condensation and the phase structure when the
backreaction is turned on.

\begin{figure}[h]
\includegraphics[width=180pt]{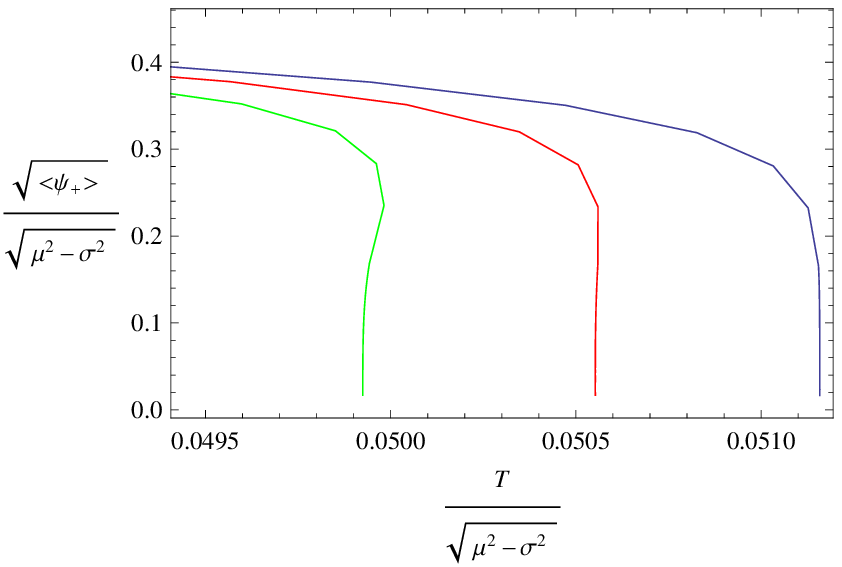}\
\includegraphics[width=180pt]{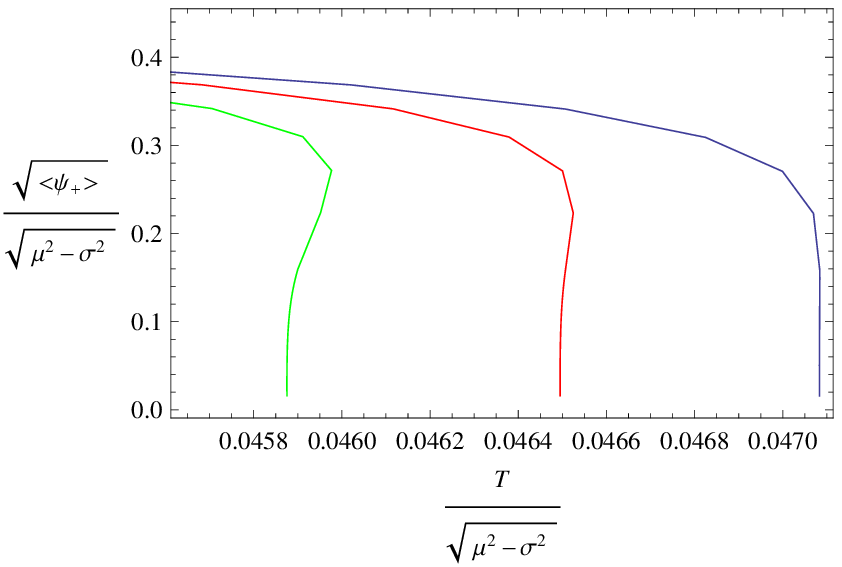}\
\caption{\label{fig5}(Color online) We plot the condensation of the
scalar operator $<O_+>$ with the change of the ratio $\sigma/\mu$.
The left panel is for choosing $\gamma=0$. Lines from right to left
are for $\frac{\sigma}{\mu}=0.26$, $\frac{\sigma}{\mu}=0.27$
\emph{and} $\frac{\sigma}{\mu}=0.28$. The right panel is for
choosing $\gamma=0.1$. Lines from right to left correspond to
$\frac{\sigma}{\mu}=0.21$, $\frac{\sigma}{\mu}=0.22$ \emph{and}
$\frac{\sigma}{\mu}=0.23$. }
\end{figure}

Let us first concentrate on the operator
$<O_+>\sim \psi_+$. In Fig.1 we plot the scalar
operator as a function of the scaled temperature.
Lines from right to left are for the increase of
the ratio $\sigma/\mu$. In the left panel we show
that our result is consistent with that observed
in \cite{C. Herzog-2,P. Basu} when we neglect the
backreaction with $\gamma=0$. We observe that
$<O+>$ does not drop to zero continuously and
becomes multivalued when $\sigma/\mu$ is over the
critical value $0.27$. This phenomenon is
attributed to the change of the phase transition
from the second order to the first order \cite{C.
Herzog-2,P. Basu}. The right panel is plotted for
choosing the backreaction $\gamma=0.1$. We find
the critical $\sigma/\mu=0.22$. Further increasing
the strength of the backreaction, the critical
value of $\sigma/\mu$ to accommodate the first
order phase transition decreases further as shown
in Table I. This tells us that with the increase
of the backreaction the first order transition
can happen even for smaller fluid velocity. This
observation is consistent with that in the
St$\ddot{u}$ckelberg mechanism where it was found
that with the increase of the backreaction the
first order phase transition can be triggered for
smaller model parameter in the
St$\ddot{u}$ckelberg mechanism \cite{ Q.Y.
Pan-1,Peng Yan}. To see clearer of the effect of
the backreaction on the phase structure, let us
look at Fig.2 where we fixed $\sigma/\mu=0.22$
with the change of the strength of the
backreaction. Again it shows that with the strong
enough backreaction, the first order phase
transition can happen.

\begin{table}
\begin{center}
\begin{tabular}{|c|c|c|c|c|}
\hline
$\gamma$ & ~~~$0$~~~ & ~~~$0.2$~~~  & ~~~$0.4$~~~ &~~ $0.6$~~~\\[1ex]
\hline
$\frac{\sigma}{\mu}$ & 0.27& 0.19 &0.16&0.15 \\
\hline
\end{tabular}
\caption{Critical ratio of
$\frac{\sigma}{\mu}$ for choosing operator
$<O_{+}>$ with various $\gamma$.}
\end{center}
\end{table}

\begin{figure}[h]
\includegraphics[width=180pt]{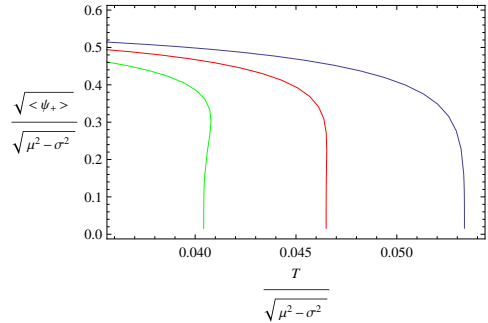}\
\caption{\label{fig5}(Color online)  We plot the scalar operator
$<O_+>$ as a function of the scaled temperature by fixing
$\frac{\sigma}{\mu}=0.22$. Lines from right to left are for choosing
$\gamma=0$, $\gamma=0.1$ \emph{and} $\gamma=0.2$.}
\end{figure}

Now let's turn to examine the behavior of the
second operator $<O_->\sim \psi_-$. When we
neglect the backreaction, the result is shown in
the left panel of Fig.3, which is consistent with
that reported in \cite{C. Herzog-2,P. Basu}. The
critical value of $\sigma/\mu$ to trigger the
first order phase transition is $0.34$. The
second operator $<O_->$ exhibits similar behavior
to that of the first operator when it comes to DC
superconductivity.

\begin{figure}[h]
\includegraphics[width=180pt]{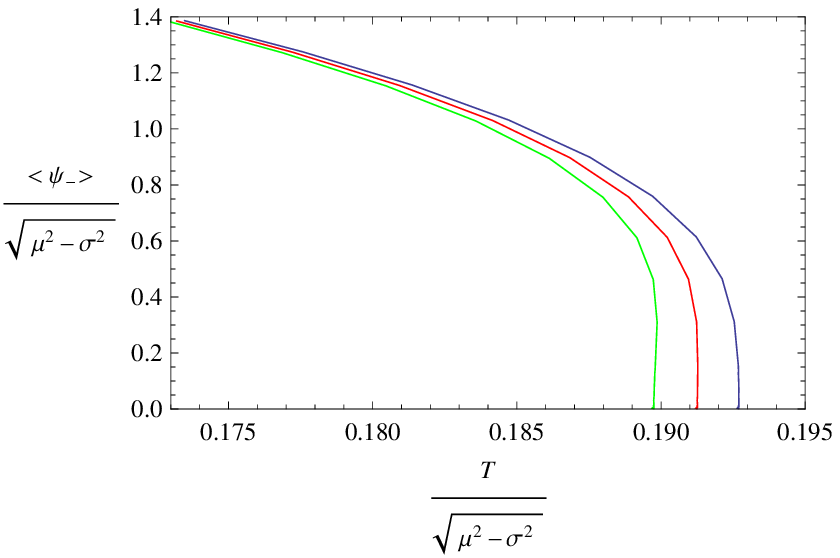}\
\includegraphics[width=180pt]{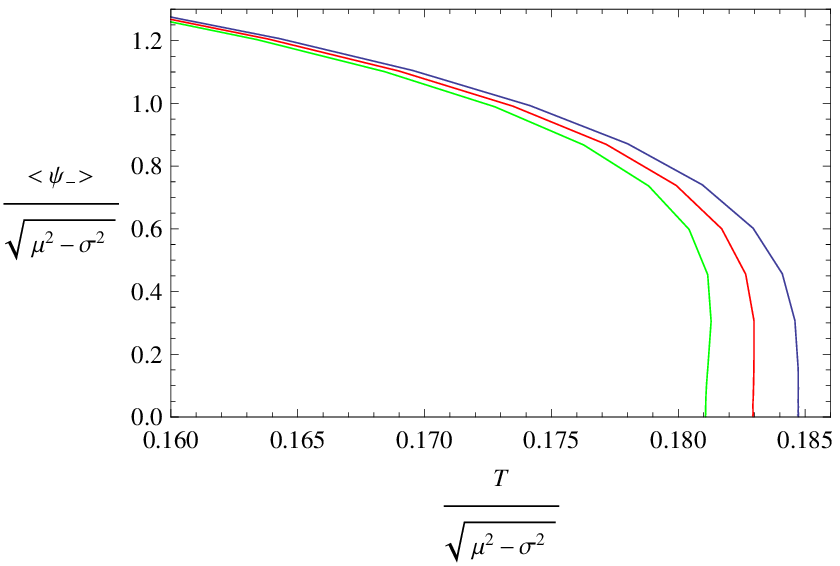}\
\caption{\label{fig5}(Color online) We plot the condensation of the
scalar operator $<O_->$ with the change of the ratio $\sigma/\mu$.
The left panel is for choosing $\gamma=0$. Lines from right to left
are for $\frac{\sigma}{\mu}=0.33$, $\frac{\sigma}{\mu}=0.34$
\emph{and} $\frac{\sigma}{\mu}=0.35$. The right panel is for
choosing $\gamma=0.1$. Lines from right to left correspond to
$\frac{\sigma}{\mu}=0.37$, $\frac{\sigma}{\mu}=0.38$ \emph{and}
$\frac{\sigma}{\mu}=0.39$.  }
\end{figure}

When we turn on the backreaction, we find the
disagreement of the behaviors of the condensation
disclosed by operator $<O_->$ from that of
$<O_+>$. With the increase of the strength of the
backreaction, the critical values of $\sigma/\mu$
to accommodate the first order phase transition
are shown in the right panel of Fig.3 and Table
II. The critical value increases when the
backreaction becomes stronger, instead of
decreasing as shown for the operator $<O_+>$.
Furthermore when we fix the ratio $\sigma/\mu$,
the first order phase transition can give way to
the second order phase transition with the
increase of the backreaction as shown in Fig.4.

\begin{table}
\begin{center}
\begin{tabular}{|c|c|c|c|c|}
\hline
$\gamma$ & ~~~$0$~~~ & ~~~$0.2$~~~  & ~~~$0.4$~~~ &~~ $0.6$~~~\\[1ex]
\hline
$\frac{\sigma}{\mu}$ & 0.34& 0.40 &0.44&0.47 \\
\hline
\end{tabular}
\caption {Critical ratio of $\frac{\sigma}{\mu}$ for $<O_{-}>$ with various $\gamma$.}
\end{center}
\end{table}

\begin{figure}[h]
\includegraphics[width=180pt]{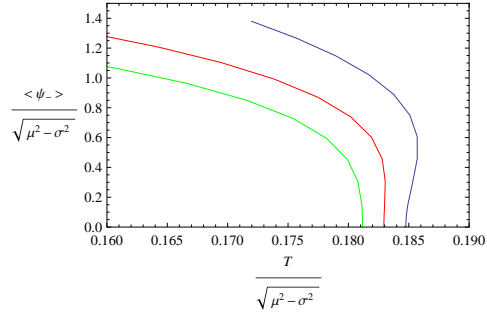}\
\caption{\label{fig5}(Color online)  We plot the scalar operator
$<O_->$ as a function of the scaled temperature by fixing
$\frac{\sigma}{\mu}=0.38$. Lines from right to left are for choosing
$\gamma=0$, $\gamma=0.1$ \emph{and} $\gamma=0.2$.}
\end{figure}

With the backreaction, the drastic different
behaviors shown in different operators in the
phase diagram  bring us the question: which
operator is physical and can reflect the real
property of the condensation in the
superfluidity.

To answer this question, let us do the rescale
and plot the figures of the condensation again.
Neglecting the backreaction, the condensation
behaviors presented by two operators are shown in
Fig.5. With the increase of the ratio
$\sigma/\mu$, the condensation gap becomes
higher, which means that the scalar hair can be
more difficult to be formed in the AdS black hole
background. Both operators $<O_+>$ and $<O_->$
present the consistent behaviors in the probe
limit. With the increase of the fluid velocity,
the critical temperature decreases. This actually
agrees with the left panels shown in Fig.1 and
Fig.3, where the temperature decreases with the
increase of the ratio $\sigma/\mu$.

\begin{figure}[h]
\includegraphics[width=180pt]{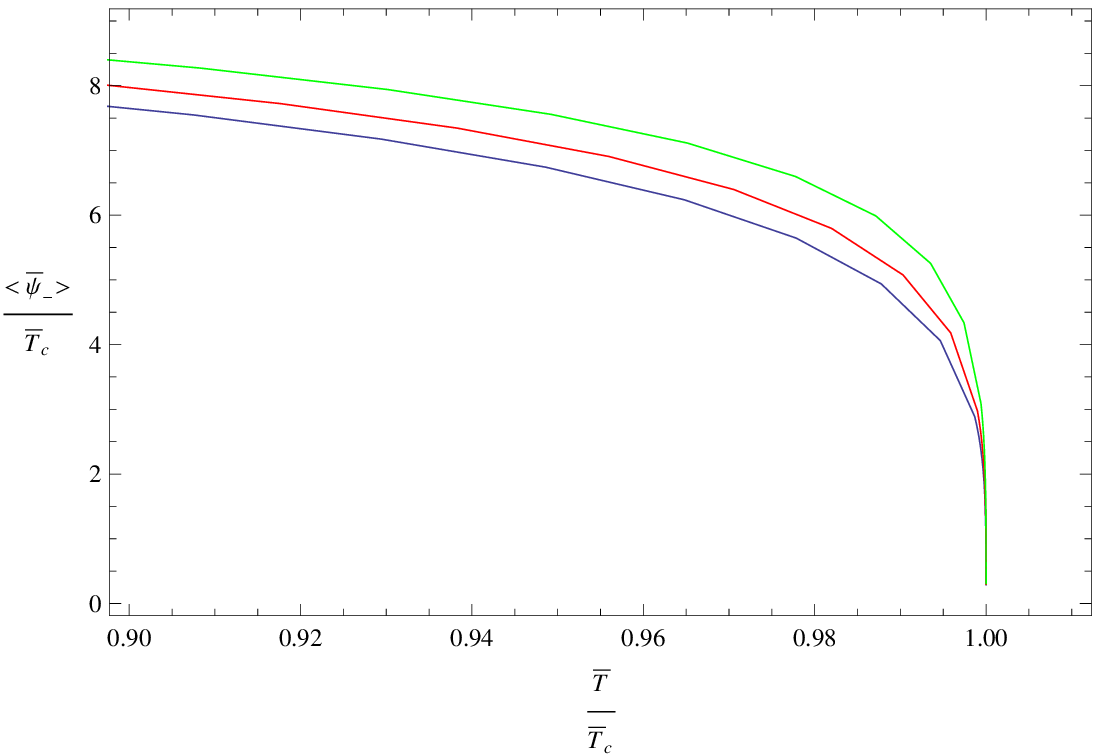}\
\includegraphics[width=180pt]{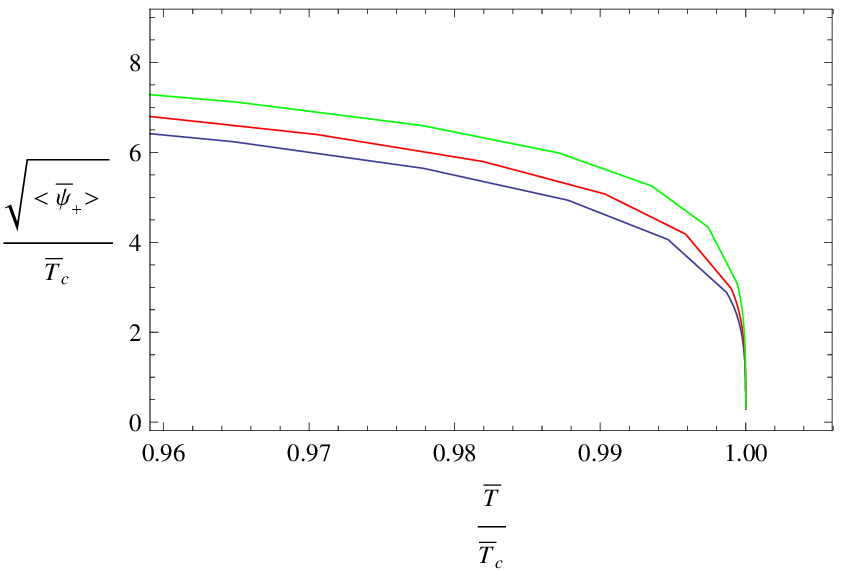}\
\caption{\label{fig5}(Color online) We plot the condensations of
operators $<O_->$ (left) and $<O_+>$ (right) in the probe limit.
Lines from top to bottom correspond to $\frac{\sigma}{\mu}=0.23 $,
$\frac{\sigma}{\mu}=0.20 $ \emph{and} $\frac{\sigma}{\mu}=0.17 $. In
the left panel the critical temperatures for lines from top to
bottom are $\overline{T}_{c}=0.204, 0.206~ \emph{and} ~0.208 $. In
the right panel the critical temperatures for  lines from top to
bottom correspond to
 $\overline{T}_{c}=0.0528,
0.0543 ~\emph{and}~ 0.0556$. }
\end{figure}

Now we replot the condensation for the system
with fully backreaction.  In Fig.6 the right
panel is for the operator $<O_+>$. We find that
with the increase of the strength of the
backreaction, the gap of the condensation becomes
higher and the critical temperature decreases.
The condensation gap marks the ease of the scalar
hair to be formed \cite{R. Gregory,Q.Y. Pan} and
its consistency with the critical temperature
indicates that the backreaction hinders the
condensation of the scalar field. This property
holds no matter the value of the fluid velocity
we choose. It is consistent with the effect of
the backreaction on the condensation when $A_x=0$
observed in \cite{Y. Brihaye}. In the left panel
we show the behavior of the condensation for the
operator $<O_->$ in the fully backreacted
geometry. We see that for the chosen
$\sigma/\mu$, when the backreaction becomes
stronger, the gap of the condensation decreases.
However the critical temperature $T_c$ decreases
as well with the increase of the backreaction.
The effect of the backreaction on the
condensation shown by $<O_->$ is different from
that when $A_x=0$. The inconsistency between the
condensation gap and the critical temperature
tells us that the operator $<O_->$ is not
appropriate to describe the condensation. Recall
that the expectation value of this operator is
related to the asymptotic behavior of the scalar
field $\sim 1/r$ at the spatial infinity, which
is the same as the operator disclosed incapable
of reflecting the correct condensation of the
scalar hair in \cite{Yunqi Liu,S. A. Hartnoll-3}.

\begin{figure}[h]
\includegraphics[width=180pt]{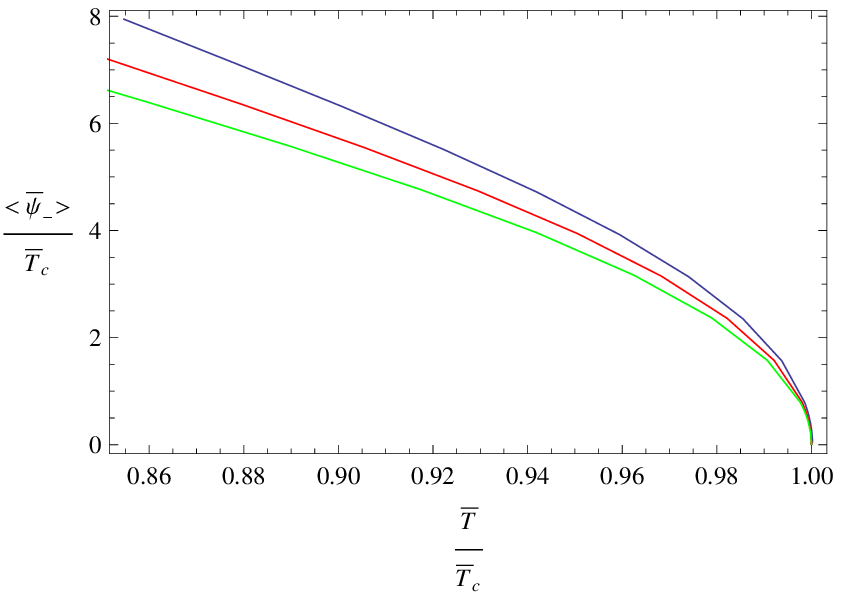}\
\includegraphics[width=180pt]{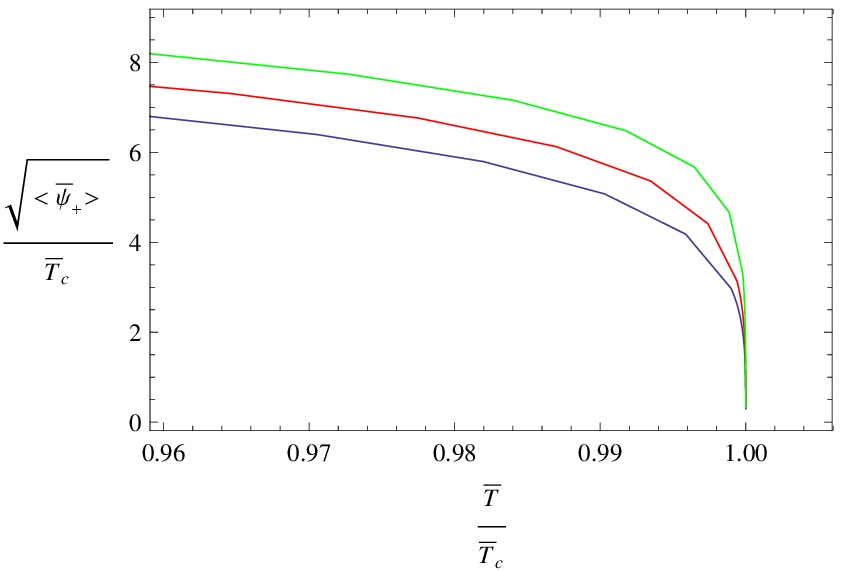}\
\caption{\label{fig5}(Color online) We plot the condensations of
operators $<O_->$ (left) and $<O_+>$ (right) with backreactions by
fixing $\frac{\sigma}{\mu}=0.20 $.  In the left panel the strength
of the backreaction for lines from top to bottom are  $\gamma=0 $,
$\gamma=0.05$ \emph{and} $\gamma=0.1 $; and the critical
temperatures read $\overline{T}_{c}=0.2060, 0.2051 ~\emph{and}~
0.2005$. In the right panel lines from top to bottom show
$\gamma=0.1$, $\gamma=0.05$ \emph{and} $\gamma=0$ with the critical
temperatures $\overline{T}_{c}=0.0476, 0.0509 ~\emph{and}~ 0.0543
$.}
\end{figure}

Besides the spatial dependence of the vector
potential described in (3), it is of interest to
examine the result by choosing other forms of the
vector potential. In the left panel of Fig.7 we
report the result by selecting
$A=\phi(r)dt+h(r)dy$. We only concentrate on the
operator $<O_+>$. The light grey regions in Fig.7
indicate the second order phase transition and
white regions are for the first order phase
transition. We see that in the probe limit, the
result is consistent with that by choosing (3).
Over the critical value of the ratio
$\sigma/\mu=0.27$, the phase transition becomes
the first order. When the backreaction is turned
on, we see the difference in choosing different
vector potential forms. For the comparison, in
the right panel of Fig.7 we plot the critical
value of $\sigma/\mu$ with the change of
backreaction $\gamma$ for choosing the vector
potential in the form of (3). In the left panel
we see that when $\sigma/\mu<0.24$, there is
always second order phase transition. When
 $0.24<\frac{\sigma}{\mu}<0.27$, with the increase of $\gamma$, the second order phase
transition can give way to the first order. For
big enough backreaction, there will be another
change of the phase structure from the first
order to the second order transition. This is
consistent with that described in cigar-shaped
solution in (3+1)-dimensional AdS spacetime
\cite{bb}. In the right panel, we observe that
when $\frac{\sigma}{\mu}<0.15$, the phase
transition is always of the second order. When
$\frac{\sigma}{\mu}$ is in the region
$0.15<\frac{\sigma}{\mu}<0.27$,  the phase
transition will change from the second order to
the first order. For the vector potential in the
form (3), we find that numerical calculation
becomes more time consuming when the backreaction
becomes stronger. In our computation till
$\gamma=0.6$ we have not observed the change of
order of  phase transition from the first to the
second as shown in the right panel.

\begin{figure}[h]
\includegraphics[width=180pt]{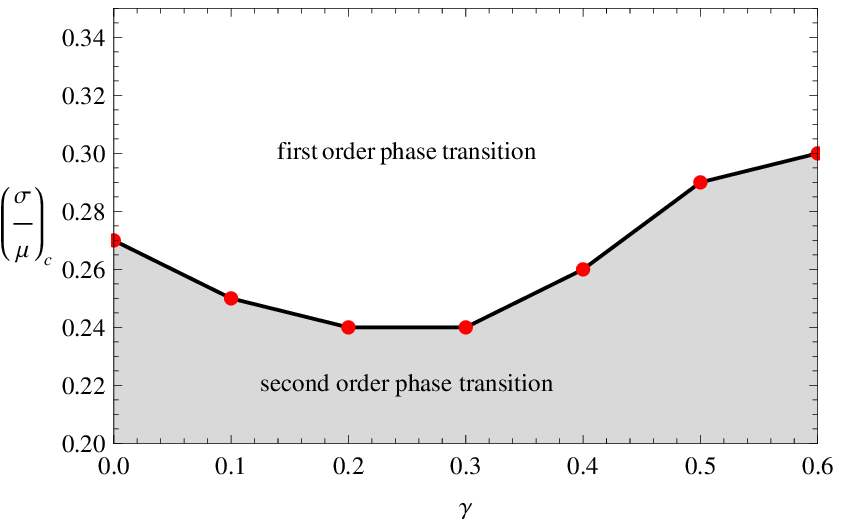}\
\includegraphics[width=180pt]{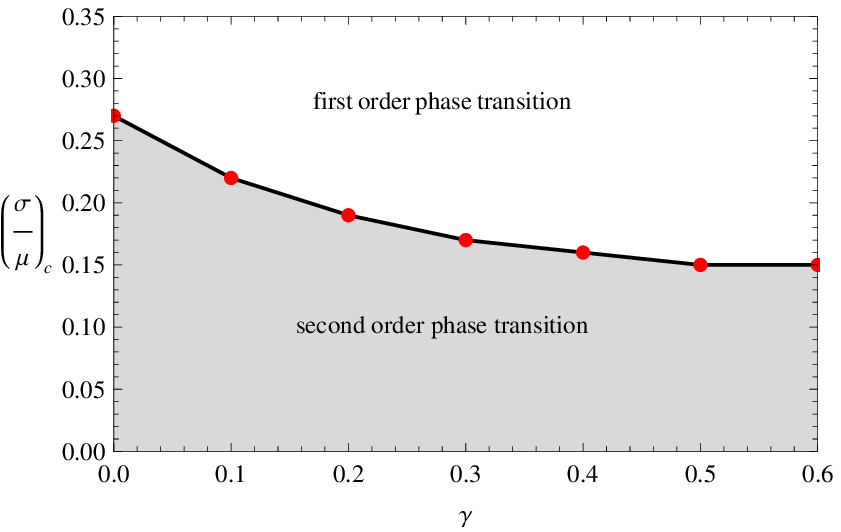}\
\caption{\label{fig5}(Color online) We plot the critical ratio
$(\frac{\sigma}{\mu})_{c}$ with backreaction parameter $\gamma$. The
left panel is related to the matter fields in the form of
$A=\phi(r)dt+h(r)dy$, while the right panel corresponds to the form
of (3). The five points from left to right correspond to $\gamma=0,
\gamma=0.1, \gamma=0.2, \gamma=0.3, \gamma=0.4, \gamma=0.5
~\emph{and}~ \gamma=0.6$ respectively and the black lines are
obtained by fitting the points in both pictures. The grey fields
represent the domain where the second order phase transition happen}
\end{figure}

\section{The condensation in AdS Soliton background}

The holographic duals to the AdS soliton in the
Maxwell field  with only nonzero vector potential
$A_t$ were studied in \cite{Q.Y. Pan,T.
Nishioka-1,G.T. Horowitz-5,A. Akhavan,ff,P.
Basu-1,Y. Brihaye,R.G. Cai,Peng Yan,Q.Y. Pan-2}.
In the probe limit, it was argued that only the
second order phase transition can happen in the
AdS soliton background \cite{Q.Y. Pan, T.
Nishioka-1,aa}. In the generalized
St$\ddot{u}$ckelberg formalism, it is possible to
change the phase transition from the second order
to the first order even in the probe limit
between the insulator and superconductor in the
AdS soliton configuration\cite{Peng Yan}.

It is of interest to generalize our discussion
above  to the bulk AdS soliton configuration by
including the spatial component of the gauge
fields with a non-trivial radial dependence.  We
will examine the scalar condensation and the
order of the phase transition between the
holographic superconductor and insulator systems.

We use the metric ansatz for the AdS soliton
\cite{Gary T}
\begin{eqnarray}\label{metric}
ds^{2}&=&-r^{2}e^{C(r)}dt^{2}+\frac{dr^{2}}{r^2B(r)}+r^{2}dx^{2}+r^2B(r)e^{D(r)}d\eta^{2}~,
\end{eqnarray}
with $B(r)$ vanishes at some radius $r_{0}$, which is the tip of the
soliton. In order to get a solution smooth at the tip we require
that $\eta$ to be periodic with a period: $\gamma=\frac{4\pi
e^{-D(r_{0})/2}}{r_{0}^2B'(r_{0})}.$

We consider solutions of the forms:
\begin{eqnarray}\label{MF}
\psi=\psi(r),~~~~~~~~~A=\phi(r)dt+h(r)d\eta,
\end{eqnarray}

With the ansatz of the spacetime, the scalar and
Maxwell equations become
\begin{eqnarray}\label{psi-3}
\psi''+(\frac{4}{r}+\frac{D'}{2}+\frac{C'}{2}+\frac{B'}{B})\psi'+(\frac{\phi^{2}}{r^{4}Be^{C}}-\frac{h^{2}}{r^{4}B^{2}e^{D}}-\frac{m^{2}}{r^{2}B})\psi=0,
\end{eqnarray}
\begin{eqnarray}\label{phi-3}
\phi''+(\frac{2}{r}+\frac{D'}{2}+\frac{B'}{B}-\frac{C'}{2})\phi'-\frac{2\psi^{2}}{r^{2}B}\phi=0,
\end{eqnarray}
\begin{eqnarray}\label{phi-3}
h''+(\frac{2}{r}+\frac{C'}{2}-\frac{D'}{2})h'-\frac{2\psi^{2}}{r^{2}B}h=0.
\end{eqnarray}
It is interesting to find that (21) and (22) are
similar to (6) and (5), respectively. This
similarity will influence the phase structure as
we will discuss below.

The nontrivial components of Einstein's equations
can be combined into
\begin{eqnarray}\label{phi-3}
C''+\frac{1}{2}C'^{2}+(\frac{4}{r}+\frac{D'}{2}+\frac{B'}{B})C'-\gamma\frac{\phi'^{2}}{r^{2}e^{C}}-\gamma\frac{2\phi^{2}\psi^{2}}{r^{4}Be^{C}}=0,
\end{eqnarray}
\begin{eqnarray}\label{phi-3}
B'(\frac{2}{r}-\frac{C'}{2})-\frac{1}{2}BD'C'+\frac{6}{r^{2}}B-\frac{6}{r^2}-\gamma\frac{h'^2}{2r^2e^{D}}+\gamma\frac{m^2\psi^2}{r^2}+\gamma
B\psi'^{2}-
\gamma\frac{\psi^2h^2}{r^4Be^{D}}+\gamma\frac{B\phi'^2}{2r^{2}e^{C}}+\gamma\frac{\phi^{2}\psi^2}{r^4e^{C}}=0,
\end{eqnarray}
\begin{eqnarray}\label{phi-3}
D'=\frac{4rC'+r^2C'^2+2r^2C''+\gamma r
(\frac{-2\phi'^{2}}{re^{C}}+4r\psi'^{2}-\frac{4\psi^2h^2}{r^3B^2e^{D}})}{r(4+rC')}.
\end{eqnarray}

In contrast to black hole, here we choose the initial boundary
condition as:

\begin{eqnarray}\label{boudary}
\phi(r)=\widetilde{\alpha}+\widetilde{\beta}(r-r_{0})+\widetilde{\gamma}(r-r_{0})^{2}+\cdot\cdot\cdot,\\
h(r)=\widetilde{a}(r-r_{0})+\widetilde{b}(r-r_{0})^{2}+\cdot\cdot\cdot.
\end{eqnarray}
The initial forms of the other fields and the
expansions of the fields at infinity are the same
as the case in back hole.

We can integrate equations numerically from the
tip $r_{0}$ out to infinity. We will concentrate
on the operator $<O_+>$ in our discussion below.
There is a critical chemical potential $\mu_{c}$,
above which the scalar hair will be formed as
reported in \cite{Tatsuma,Gary T}. Neglecting the
backreaction, $\gamma=0$, we observe the behavior
of the condensation in the left panel of Fig.8.
With the increase of the ratio $\sigma/\mu$, the
condensation gap increases and the critical
chemical potential $\mu_c$ decreases. This is
consistent with the situation we observed in the
AdS black hole where $\mu_c$ in the soliton plays
the similar role to $T_c$ in the black hole.
However in the soliton configuration we see that
the operator $<O_+>$ drops to zero continuously
and the phase transition is always of the second
order no matter how big the ratio $\sigma/\mu$
is. This is drastically different from the result
seen in the AdS black hole case. For the chosen
value of $\frac{\sigma}{\mu}$, the dependence of
the condensation on the backreaction is shown in
the right panel of Fig.8. The second order phase
transition is still kept when the backreaction is
taken into account.
\begin{figure}[h]
\includegraphics[width=180pt]{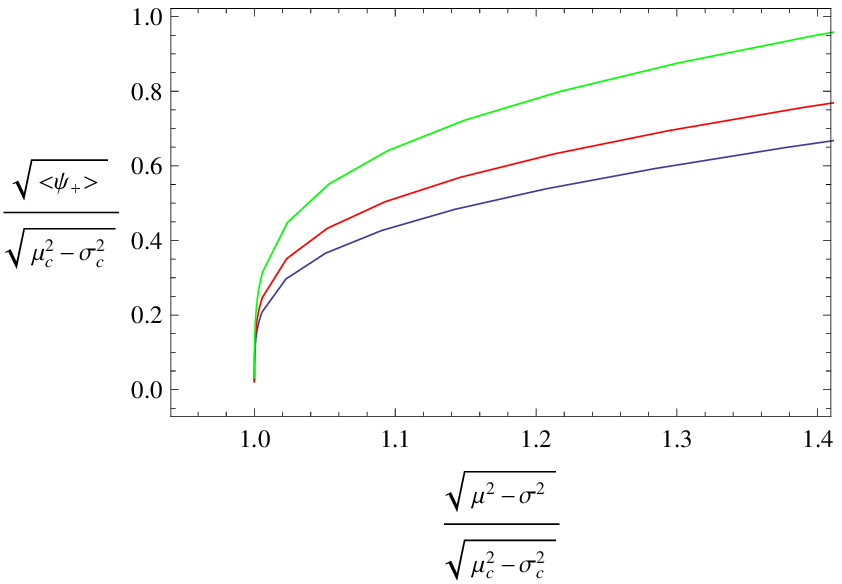}\
\includegraphics[width=180pt]{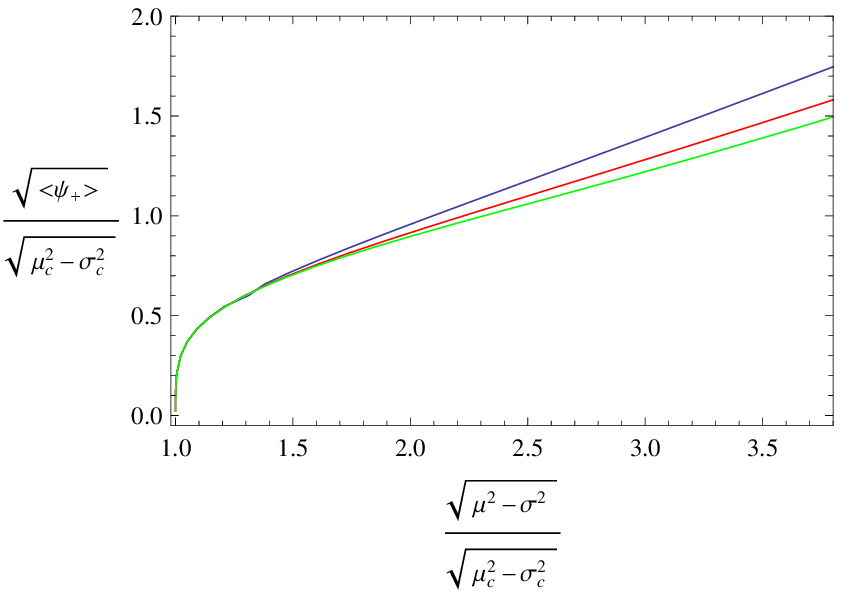}\
\caption{\label{fig5}(Color online) (Left) We plot the condensation
with the change of the ratio $\sigma/\mu$ in the probe limit. Lines
from top to bottom correspond to $\frac{\sigma}{\mu}=0.8$,
$\frac{\sigma}{\mu}=0.6$ \emph{and} $\frac{\sigma}{\mu}=0$ with the
critical values of the chemical potential
$\sqrt{\mu_{c}^{2}-\sigma_{c}^{2}}=1.0823, 1.4123 ~\emph{and}
~1.7182$ respectively. (Right) We fix the ratio $\sigma/\mu=0.2$.
Lines from top to bottom correspond to $\gamma=0$, $\gamma=0.6$
\emph{and} $\gamma=0.8$ with the critical chemical potential
$\sqrt{\mu_{c}^{2}-\sigma_{c}^{2}}=1.6885, 1.6911~ \emph{and}
~1.6920$ respectively.}
\end{figure}

To understand the reason why the spatial
component $A_x$ for the vector potential plays
different role in the phase diagrams in the AdS
black hole and AdS soliton backgrounds, we plot
$\phi[u]$ and $h[u]$, where $u=\frac{1}{r}$, in
both backgrounds in Fig.9 around the point where
the phase transitions begin to happen.

It is clear that in the AdS soliton background
the spatial component of the gauge field  behaves
very different from that in the AdS black hole
configuration. The $h[u]$ in the AdS soliton
behaves similar to $\phi[u]$ in the black hole,
thus similar to the time component $A_t$ in the
AdS black hole,  the spatial component in the AdS
soliton background  cannot bring first order
phase transition. On the other hand, $\phi[u]$ in
the soliton behaves like $h[u]$ in the black hole
background. For this reason, it was observed that
only with the time component of the vector
potential, the first order phase transition can
happen in the five-dimensional AdS soliton
background with sufficient backreaction
\cite{Gary T}, where the charge density and the
condensate display sharp discontinuity at the
critical of $\mu$. In the four-dimensional AdS
soliton background, if we just consider the time
dependent of the vector potential, the sharp
discontinuity phenomenon in the condensate
appears when $\gamma=0.87$, which indicates the
happening of the first order phase transition.
The numerical findings is consistent by comparing
(21)(22) and (5)(6).
\begin{figure}[h]
\includegraphics[width=180pt]{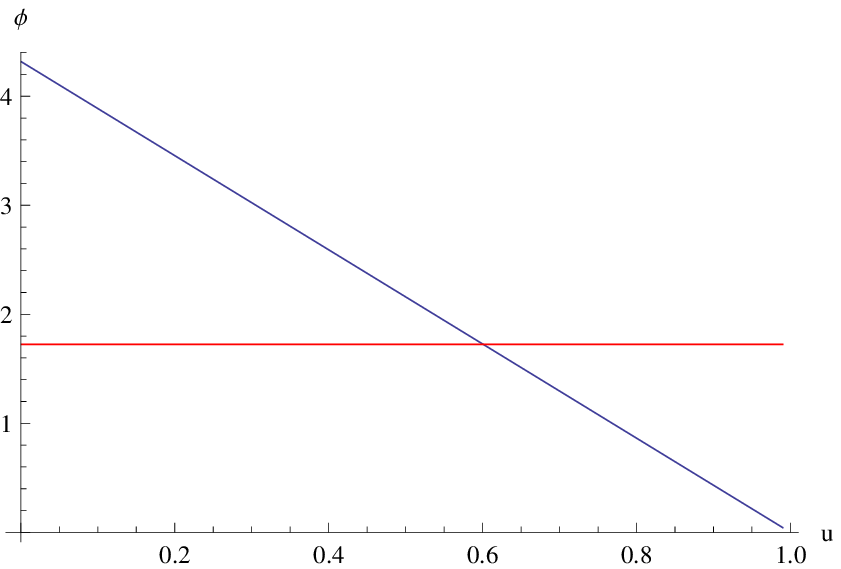}\
\includegraphics[width=180pt]{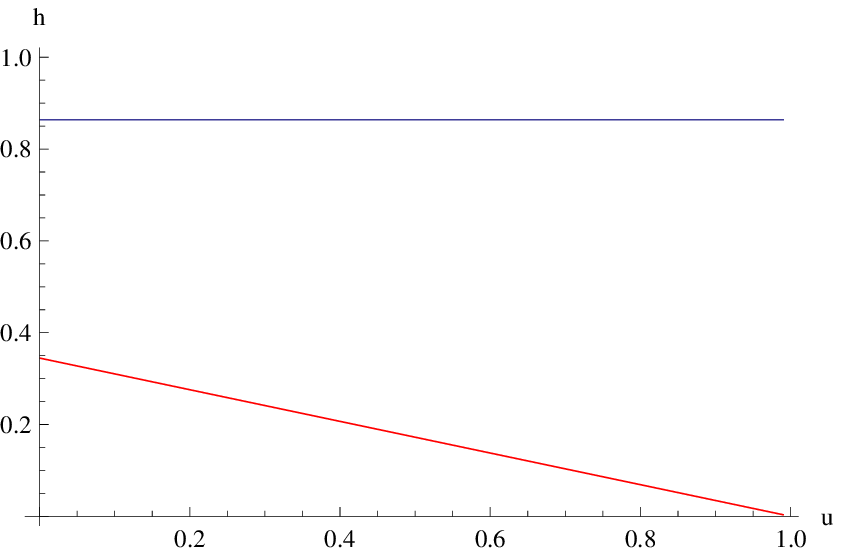}\
\caption{\label{fig5}(Color online) We plot the
behaviors of the time component and spatial
component of vector potentials in both the AdS
black hole and AdS soliton backgrounds. We choose
$\gamma=0.1$, $\frac{\sigma}{\mu}=0.2$ and
$\psi[r_{h}]=\psi[r_{0}]=1/1000$ in plotting
these figures. The blue line and red line
correspond to the cases in AdS black hole and AdS
soliton, respectively.}
\end{figure}

\section{Conclusions and discussions}

In this work we have studied a static solution to
the system with a charged scalar field coupled to
the AdS black hole, which in the dual field
theory corresponds to a static current flowing in
a superconducting fluid. We have considered the
fully back reacted geometry and found the
influence of the backreaction on the phase
structure. With the backreaction, the first order
phase transition can happen for smaller fluid
velocity.

Different from that observed in the probe limit \cite{C. Herzog-2,P.
Basu-1}, when we turned on the backreaction, we found that the
condensation behaviors and phase structures reflected from
expectation values of different operators are different. This
phenomenon was also observed in the study of the effect of the
backreaction in the holographic superconductor in \cite{S. A.
Hartnoll-3,Yunqi Liu}. Considering the gap of the condensation which
marks the ease of the scalar hair to be formed \cite{R. Gregory,Q.Y.
Pan} and its correspondence to the the critical temperature, we
concluded that the operator associated with the asymptotic behavior
$\sim 1/r$ at the spatial infinity is not appropriate to describe
the condensation. This observation is consistent with the findings
in \cite{S. A. Hartnoll-3,Yunqi Liu}.

We have further generalized our discussion to the
AdS soliton and examined the spatial component of
the gauge field on the phase structure in the AdS
soliton background. Different from the AdS black
hole background, we found that the spatial
component of the vector potential cannot bring
the first order phase transition in the soliton
background. Instead, the first order phase
transition can be brought by the time component
of the vector potential in the AdS soliton
configuration when the backreaction is strong
enough as observed in \cite{Gary T,Peng Yan}. We
have analyzed the reason behind this difference
between the AdS black hole and AdS soliton
backgrounds.

\begin{acknowledgments}
This work has been supported partially by the
 NNSF of China and the Shanghai Science and
Technology Commission under the grant
11DZ2260700.

\end{acknowledgments}

\end{document}